\documentclass[sigconf]{acmart}
\usepackage{multirow}
\usepackage{makecell}
\usepackage{subfigure}
\AtBeginDocument{%
  \providecommand\BibTeX{{%
    \normalfont B\kern-0.5em{\scshape i\kern-0.25em b}\kern-0.8em\TeX}}}

\acmSubmissionID{}

\begin{document}

\title{GhostVec: A New Threat to Speaker Privacy of End-to-End Speech Recognition System}

\author{Xiaojiao Chen}
\affiliation{%
  \institution{Xinjiang University}
  \city{Urumqi}
  \country{China}
}\email{xiaojiaoch@163.com}

\author{Sheng Li}
%\authornote{This work was supported by JSPS KAKENHI Grant Numbers JP23K11227, JP22H03595, JP23H03402.}
\affiliation{%
  \institution{NICT}
  \city{Kyoto}
  \country{Japan}}
\email{sheng.li@nict.go.jp}

\author{Jiyi Li}
\affiliation{%
  \institution{University of Yamanashi}
  \city{Kofu}
  \country{Japan}
}\email{jyli@yamanashi.ac.jp}

\author{Yang Cao}
\affiliation{%
  \institution{Hokkaido University}
  \city{Sapporo}
  \country{Japan}}
\email{yang@ist.hokudai.ac.jp}

\author{Hao Huang}
\affiliation{%
 \institution{Xinjiang University}
 \streetaddress{Rono-Hills}
  \city{Urumqi}
  \country{China}}
\email{hwanghao@gmail.com}

\author{Liang He}
\affiliation{%
  \institution{Tsinghua University}
  \city{Beijing}
  \country{China}}
\email{heliang@tsinghua.edu.cn}

\renewcommand{\shortauthors}{Chen, et al.}

\begin{abstract}
Speaker adaptation systems face privacy concerns, for such systems are trained on private datasets and often overfitting. This paper demonstrates that an attacker can extract speaker information by querying speaker-adapted speech recognition (ASR) systems. We focus on the speaker information of a transformer-based ASR and propose GhostVec, a simple and efficient attack method to extract the speaker information from an encoder-decoder-based ASR system without any external speaker verification system or natural human voice as a reference. To make our results quantitative, we pre-process GhostVec using singular value decomposition (SVD) and synthesize it into waveform. Experiment results show that the synthesized audio of GhostVec reaches 10.83\% EER and 0.47 minDCF with target speakers, which suggests the effectiveness of the proposed method. We hope the preliminary discovery in this study to catalyze future speech recognition research on privacy-preserving topics.
\end{abstract}






%\renewcommand\footnotetextcopyrightpermission[1]{}
\settopmatter{printacmref=false}
\maketitle

\section{Introduction}
Automatic speech recognition (ASR) is a key technology in many speech-based applications, e.g., mobile communication devices and personal voice assistants, which typically require users to send their speech to the system for recognized text. Thanks to the development of deep learning and machine learning, ASR systems \cite{graves2006connectionist,gulati2020conformer,dong2018speech,schneider2019wav2vec,radford2022robust} have received a tremendous amount of attention and show excellent advantages over traditional ASR systems.

There is a problem that the performance of ASR can still degrade rapidly when their conditions of use differ from the training data. Many adaptive algorithms (speaker adaptation \cite{meng2019adversarial,shi2020h,li2018speaker}, domain adaptation \cite{hosseini2018augmented,meng2019domain}, accent adaptation \cite{yang2018joint,turan2020achieving}, etc.) are used to alleviate the mismatch between the training data and test data. 
Speaker adaption, which adapts the system to a target speaker, is also one of the most popular forms of adaptation. Speaker adaptation attracts the attention of many researchers: it can explicitly model speaker characteristics and the speech context \cite{bell2020adaptation}. 
Some researchers \cite{saon2013speaker} directly extract speaker embeddings such as i-vectors in a manner independent of the model and concatenate them with acoustic input features. 
Instead of directly using speaker embedding as a speaker feature, some studies have proposed extracting low-dimensional embeddings from bottleneck layers in neural network models trained to distinguish between speakers \cite{shi2020h} or across multiple layers followed by dimensionality reduction in a separate layer. Unlike the above speaker embedding approaches, some works do not use speaker information explicitly for ASR. 
\cite{tang2016multi,kanda2020joint} used multi-task learning (MTL) to unify the training of transcribing speech and identifying speakers simultaneously by sharing the same speech feature extraction layers. Adversarial learning (AL) adopts a similar architecture to MTL but learns a speaker-invariant model to be more generalized to new speakers by reducing the effects of speaker variability \cite{meng2019adversarial,sun2018domain}. 
Those methods, however, face the challenge of overfitting to targets seen in the adaptation data. 

\begin{figure*}[htb]
    \centering
	\includegraphics[scale=0.8]{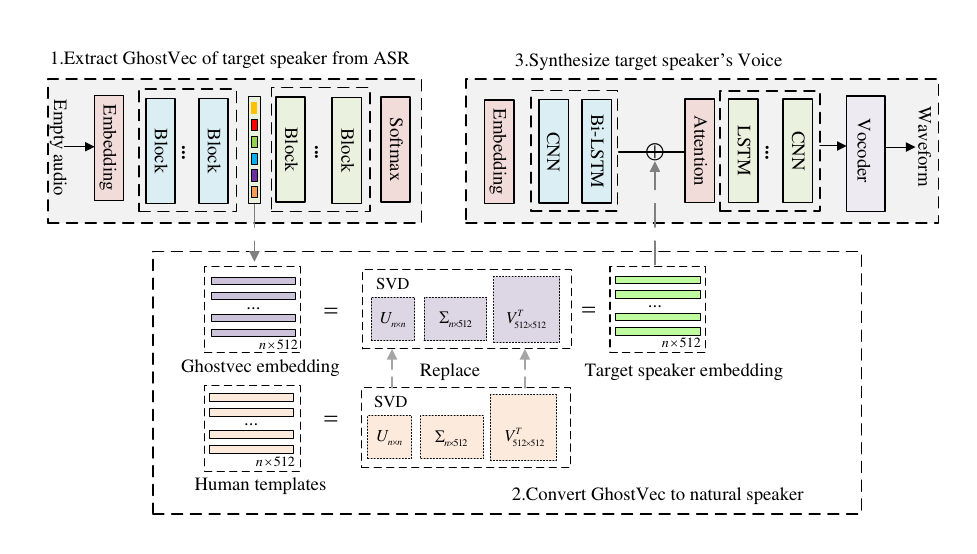}
	\caption{The proposed method's flowchart consists of extracting GhostVec, converting GhostVec to a natural speaker, and synthesizing the target speaker's voice. }
	\label{fig:svd}
\end{figure*}

Recently, machine learning models have gained notoriety for exposing information about their training data, which can cause privacy leaks.
Researchers \cite{yeom2018privacy} view overfitting as a sufficient condition for privacy information leakage, and many attacks work by exploiting overfitting \cite{shokri2017membership,carlini2021extracting}. An attacker \cite{shokri2017membership} can apply a member inference attack to predict the presence of some specific examples in the training data. 
In \cite{carlini2021extracting}, the authors also illustrate how large language models can be prodded to disgorge sensitive, personally identifying information picked up from their training data. 
As speech recognition algorithms increasingly have free access to everyone, attackers have more opportunities to extract private training data. Adversarial example \cite{goodfellow2014explaining} is one of the well-known attack methods, and it is usually generated by adding some small perturbation to the example. 
Since almost all of the state-of-the-art ASR systems are based on the transformer-based model, it is necessary to investigate the leakage of speaker information in this model. This paper's goal and our findings might serve as a warning in privacy-preserving speaker-adaptive ASR systems.

\begin{figure}[t]
    \centering
	\includegraphics[scale=0.46]{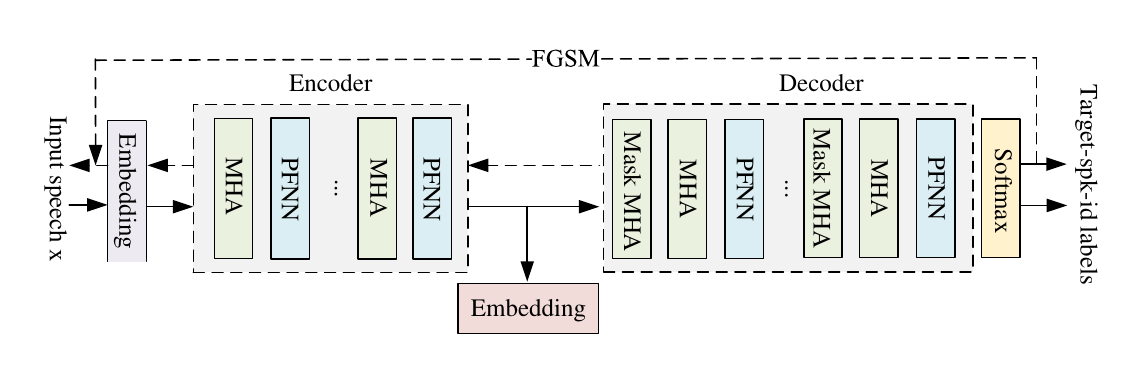}
	\caption{Extracting GhostVec of target speaker from a pre-trained model. Multi-head self-attention (MHA) and position-wise feed-forward networks (PFFN). The input is background noise, and the speaker-id is the target speaker. The embedding output of the encoder is GhostVec.}
	\label{fig:gv}
\end{figure}

This paper focuses on a speaker-adaptive transformer-based ASR system, which includes massive speaker and acoustic information. It investigates the privacy leakage of speaker information from speaker-adapted ASR.
Specifically, we propose GhostVec extract the target speaker's voice from the existing ASR model by adversarial example. 
We process GhostVec to synthesize the target speaker's voice to make the speaker information further distinguishable.

The main contributions can be summarized as follows.
(1) We raise the concern of GhostVec, the adversarial speaker embedding. It can lead to speaker information leakage from speaker-adapted End-to-End ASR systems. 
(2) We also show a method to synthesize the targeted speaker voice from GhostVec.

\section{Proposed Methods}
This work demonstrates that extracting speaker information from a trained speaker-adaptive ASR model is possible. This section discusses the proposed method, in which we first generate a target speaker's GhostVec and further convert it to the target speaker's voice. Fig.\ref{fig:svd} illustrates the flowchart of the proposed method. The following subsections are detailed descriptions.

\begin{figure}[t]
\centering
\includegraphics[scale=0.6]{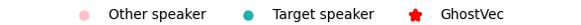}
\includegraphics[scale=0.45]{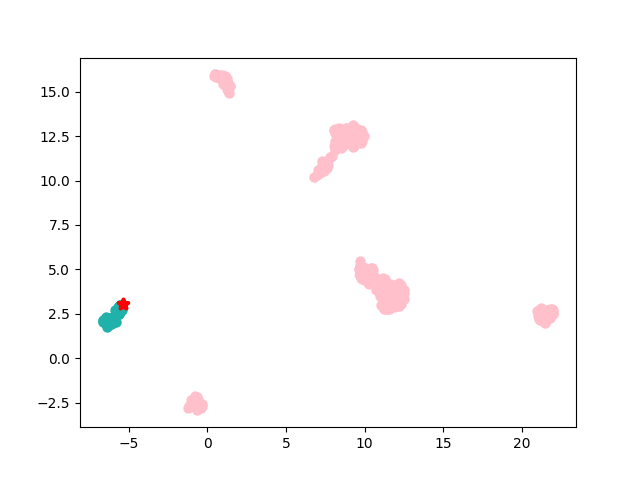}
\caption{UMAP \cite{mcinnes2018umap} similarity between target speakers' embeddings and GhostVecs. There are six speakers in this figure.}
\label{fig:umap}
\end{figure}

\begin{table}[t]
  \begin{center}
  \caption{Subjective evaluations (95\% confidence interval).}
  \label{tab:subjective}
     \begin{tabular}{l|cc}
      \toprule
    %   & Naturalness & Similarity  \\
     methods & MOS & DMOS   \\
      %\hline
      \midrule
      Genuine speaker & 3.91& 3.84 \\
      %\hline
      SVD-modified GhostVec (Our) & 3.73 &2.78\\
      \bottomrule
     \end{tabular}
  \end{center}
\end{table}

\begin{table}[t] \footnotesize
  \caption{Objective evaluation. Target, GhostVec, and Proposed (SVD-modified GhostVec), respectively, mean the speaker characteristic of the synthesized audio comes from the target speaker, GhostVec, and SVD-modified GhostVec speaker embedding.} 
  \label{tab:cosine}
    %\begin{threeparttable}
    \begin{center}
    \begin{tabular}{cc|cccc}
      \toprule
      \multirow{2}{*}{Enrolls} &  \multirow{2}{*}{Trails} & \multirow{2}{*}{EER\% \cite{Snyder2018x}} & \multirow{2}{*}{minDCF} & \multicolumn{2}{c}{$C_{llr}$} 
      \\
       %\cline{5-6}
      &  & & & min &act
     \\
      %\hline
      \midrule
     Target & Target & 1.50 & 0.32 & 0.07 &0.71\\
      %\hline
        Target & GhostVec  & 52.27 & 1.00 & 0.99& 143.87\\ 
      %\hline
      Target& Our  & 10.83 & 0.46 &0.34 & 42.22 \\
      \bottomrule
    \end{tabular}
    \end{center}
\end{table}

\subsection{Extract GhostVec of the Target Speaker}
\label{ghostvecextract}

A speaker-adaptive ASR model, which adapts the system to a target speaker, can explicitly model speaker characteristics and the speech context \cite{bell2020adaptation}. In this paper, as shown in Fig.\ref{fig:gv}, we extract the information from this trained transformer-based speaker-adaptive ASR network, which includes two parts: an encoder and a decoder. The encoder is responsible for encoding the input speech feature sequence $\mathbf{x}$, and the decoder predicts the output sequence $y_{1:l-1}$ according to the decoding information from the encoder output ${\mathbf{h^E}}$.
In previous studies\cite{huang2019black,wang2021delving}, random noise was used to learn the model information for success adversarial attacks. 
To eliminate the dependence on speech, we adopt empty audio, aka background noise, as the input to extract the speaker's voice in this paper.

To gain the speaker information from the model, the concept of adversarial example is adapted to find information by extracting speaker embedding ${\mathbf{h^E}}$. We add small and purposed adversarial perturbation $\delta$ to the empty input, and $\delta$ is trained to find the speaker information in this speaker-adaptive ASR model. The output of encoder ${\mathbf{h^E}}$ represents the speaker information. We call the embedding ${\mathbf{h^E}}$ that contains the speaker's information as GhostVec. As shown in Fig.\ref{fig:gv}, the process of extracting GhostVec is 
%\begin{eqnarray}\small
$\textbf{h}^{E} = { \rm Encoder}({\textbf x}+\delta)$,
$\hat{\textbf{h}}_{l}^{D} = {\rm Decoder}(\textbf{h}^{E}, \hat{y}_{1:l-1})$,
$P(y_l| \hat{y}_{1:l-1}, \textbf{h}^{E}) =  {\rm softmax}(\hat{\textbf{h}}_{l}^{D})$,
%\end{eqnarray}
where ${\textbf x}$ is the input feature of speech, $y_{1:l}$ is the recognition result of the target speaker, which contains the speaker-id; the likelihood of each output token $y_{l}$ is given by \textup{softmax}, $y_1$ is the speaker id of the target speaker; and $\delta$ is the generated adversarial perturbation by model. 

To extract the speaker information ${\mathbf{h^E}}$, $\delta$ is optimized to help empty input $x$ to obtain the speaker information from the model. In this paper, $\delta$ is updated by the fast gradient sign method (FGSM)\cite{szegedy2013intriguing} and is formulated as:
    $\delta = \epsilon \cdot {\rm sign}(\nabla _{\textbf{x}}J(\theta,\textbf{x},\textbf{y}))$,
where $\theta$ is the ASR model's parameter and is frozen in this paper; $\epsilon$ is a hyperparameter to constrain $\delta$ with a small value; $\textbf{y}$ satisfies $\textbf{y}=f(\textbf{x})$ and is the correct output of $\textbf{x}$; and $J(\theta,\textbf{x},\textbf{y})$ is the loss function used in the trained model. 
Considering that speech content is not the aim of this paper, we only focus on the result of speaker id $y_1$. So, extracting the speaker information has a target speaker id, and $\delta$ is optimized until the output $y_1$ is a target speaker id. Embeddings from the same speaker are close in the embedding vector space, and we observe GhostVec in low dimensions. As shown in Figure \ref{fig:umap}, we show the clustering effect of GhostVec with six speakers in low dimensions. GhostVec can be clustered into the same class with the target speaker in low dimensions and distanced from the non-target speaker. We think the speaker information has been extracted. 

\subsection{Converting GhostVec to Waveform} 
GostVec has a specific and detailed meaning for the model, but humans cannot intuitively comprehend it from the embedding perspective. The authors in \cite{wu2022adversarial} prove that the vocoder’s preprocessing will not affect the ASV scores of genuine samples too much. To further make our results quantitative, we utilize the text-to-speech system (TTS) to convert GhostVec to a waveform, which is a human-comprehensible form. The TTS model synthesizes waveforms given text, and multi-speaker TTS can generate natural speech for various speakers by speaker embedding. In particular, given conditional text $c_{text}$, the TTS model aims to build the mapping $G(\cdot)$, viz., $z$ = $G(\mathbf{X}, c_{text})$, where $\mathbf{X}$ is the speaker embedding that controls the audio styles with the speaker identity, and $z$ is the synthesized waveform, which is the output of the TTS model. Therefore, different speaker embeddings $\mathbf{X}$ have different representations of the synthesized speech. 

GhostVec is nonrobust and adversarial \cite{ilyas2019adversarial}. We adopt SVD to preprocess GhostVec to a robust embedding. SVD can find the essential dimension of a matrix. We expect audio with real human intelligibility and naturalness by modifying GhostVec's decomposed matrices.
Generally, for an arbitrary matrix $ {\mathbf{X}}_{M \times N}$, SVD can be expressed as: $\mathbf{X} = \mathbf{U \Sigma V^\top}$, where $\mathbf{U}_{M \times M}$ and $\mathbf{V}_{N \times N}$ are the orthogonal matrices, $\mathbf{\Sigma}$ is a diagonal matrix, so that $\mathbf{\Sigma}$ is rectangular with the same dimensions as  ${\mathbf{X}}$. The diagonal entries of $\mathbf{\Sigma}$ that is $\mathbf{\Sigma}_{ii} = \sigma_i$, can be arranged to be nonnegative and in order of decreasing magnitude. We interpreted $\mathbf{\Sigma}$ as low-dimensional speaker information directly reflected in speaker clustering (as shown in Fig. \ref{fig:umap}). $\mathbf{U}$ and $\mathbf{V}$ reflect the intrinsic characteristics of human voices or adversarial audios.

As shown in Fig. \ref{fig:svd}, we replaced the $\mathbf{U}$ and $\mathbf{V}$ of the GhostVec with the $\mathbf{U}$ and $\mathbf{V}$ from template human voices, which is a nearest human template of target speaker in the speaker embedding spaces. In this paper, we truncate the GhostVecs for the same target speaker with different utterances (100 sentences) to organize a matrix, and then we use SVD to decompose this matrix. The distance between embeddings is measured by cosine distance.

\section{Experiment Setup}
\noindent\textbf{Datasets}: All datasets utilized in the ASR model were based on the 100 hours of Librispeech dataset (train-clean-100) \cite{librispeech2015}. The input feature was 120-dimensional log Mel-filterbank (40-dim  static, +$\Delta$, and +$\Delta\Delta$). The synthesizer of multi-speaker TTS used 100 hours of train-clean-100, and the vocoder was trained using VCTK \cite{vctk} datasets. Librispeech TTS train-clean-500 is used as template human embedding during pre-processing GhostVec. 

\begin{figure}[t]
\centering
\includegraphics[scale=0.30]{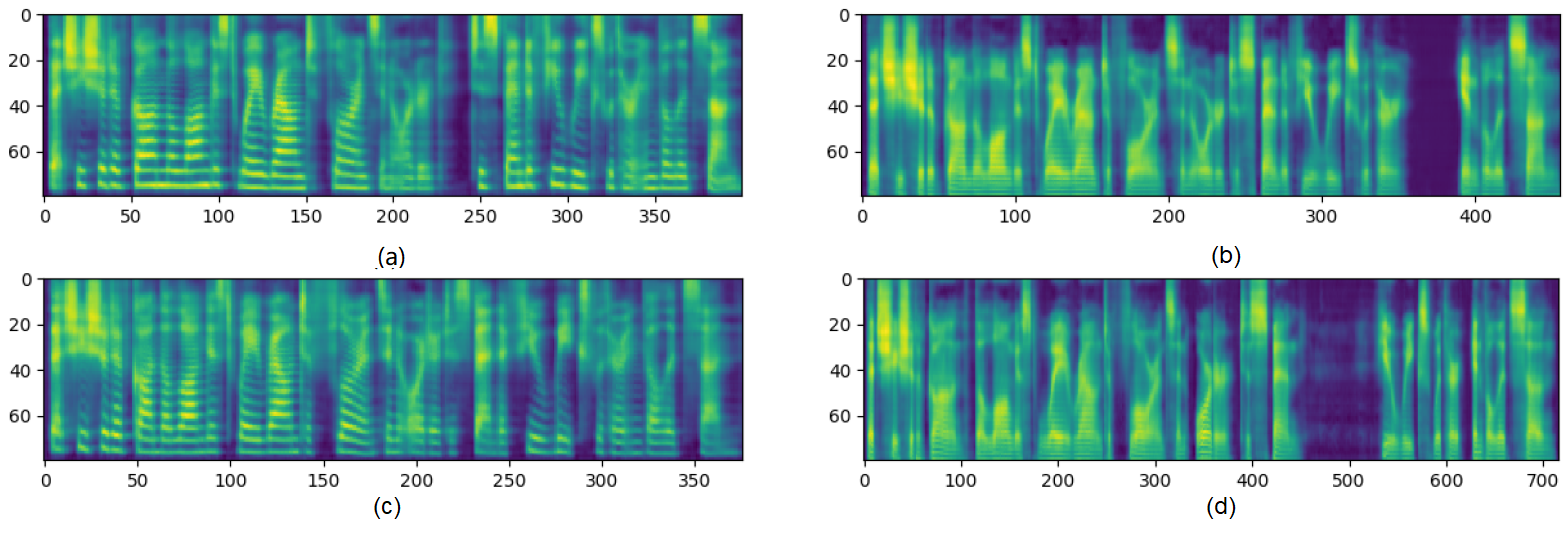}
\caption{The Comparison of mel-spectrum different embedding (the same text). 
%\footnotesize
(a) Genuine female, embedding predicted Mel-Spectrogram, (b) Genuine male, embedding predicted Mel-Spectrogram, (c) Female, SVD-modified GhostVec predicted Mel-Spectrogram, (d) Male, SVD-modified GhostVec predicted Mel-Spectrogram.
}
\label{fig:mel}
\end{figure}

\begin{table}[t]
  \begin{center}
  \caption{Objective evaluation with CER\%.}
  \label{tab:cer}
    \begin{tabular}{l|c}
      \toprule
    methods  & CER\% \\
      %\hline
      \midrule
      Baseline \cite{lishengIS-share} & 9.80\\
      %\hline  
      SVD-modified GhostVec (Our) & 15.42  \\
      \bottomrule
    \end{tabular}
  \end{center}
\end{table}

\noindent\textbf{ASR model setting:} We adopt the transformer-based speaker-adapted speech recognition model (ASR$_{spk}$). In this paper, the ASR$_{spk}$ model required for embedding extraction is trained on the LibriSpeech train-clean-100 but based on the multitask training method following \cite{lishengIS-share,attisall} with the speaker-id and label. The character error rate (CER\%) was approximately 9.0\%, which is our proposed method's baseline. The speaker-id was explicitly added as the label during training \cite{lishengIS-share,attisall}. The training labels are organized as ``$<$SOS$>$ $<$speaker-id$>$ labels $<$EOS$>$''. We augment target speaker IDs to ground truth labels in training GhostVec.

\noindent\textbf{TTS system setting:} Synthesizing the target speaker's voice is based on an End-to-End multispeaker TTS system \cite{liu2021exploring}. This system extracts speech embeddings from an ASR encoder\cite{lishengIS-share} to improve the multispeaker TTS quality, especially for speech naturalness.
We trained the synthesizer and vocoder separately. For the synthesizer, we train based on the LibriSpeech train-clean-100 and embeddings from the ASR$_{spk}$ model \cite{lishengIS-share}. We trained the vocoder using VCTK.

\section{Experiment Results}

\subsection{The quality of synthesized audio}

\subsubsection{Subjective evaluations}:
We first show that synthesized speech from our proposed method is natural and understandable. In this paper, we adopt the mean opinion score (MOS) for speech naturalness and the differential MOS (DMOS) \cite{kinnunen2018spoofing} for similarity. Specifically, we invited 21 participants who used headphones for listening tests. For the naturalness evaluation, all listeners completed 120 audio tasks \footnote{Audio samples are available at https://demoghostvec.github.io}. Additionally, all listeners completed 60 pairs of audio tasks for the similarity evaluation. The reference sentences were synthesized using 12 speakers (6 male and 6 female). Table \ref{tab:subjective} shows the
average MOS and DMOS scores with 95 \% confidence intervals. We use the genuine waveform of the target speaker as the baseline. As shown in Table \ref{tab:subjective}, the MOS of the proposed SVD-modified GhostVec has expected performance in speech naturalness. Furthermore, there is still some gap in DMOS scores, and the difference between the two groups deals with the adversarial factors in the GhostVec. We must remove the adversarial factors of GhostVec for higher-quality TTS.

\subsubsection{Objective Evaluation}
\label{objection}
As for the actual similarity measurement of synthesized audio and reference audio, we use three evaluation metrics to report the system performance: Equal Error Rate (EER), minimum Detection Cost Function (minDCF) with ptarget= 0.01 and the log-likelihood ratio (LLR) based costs $C_{llr}$, which can be decomposed into a discrimination loss ($C_{llr}^{min}$) and a calibration loss($C_{llr}^{act}$). A total of 120 sentences generated by the proposed method and the same number of audios from the different target GhostVec are also provided. Moreover, the value of EER\% from speaker verification (SV)  \cite{Snyder2018x} was used to evaluate similarity as the objective evaluation of the proposed method in Table \ref{tab:cosine}. The results in Table \ref{tab:cosine} show that our synthesized waveform effectively demonstrates target speaker characteristics. The proposed SVD-modified GhostVec method achieves better performance, effectively surpassing the GhostVec. The results suggest that the speaker's identity has been hidden, while the GhostVec is directly conveyed to the TTS system. The proposed SVD-modified GhostVec method successfully represents the target speaker identity.

\subsection{Automatic Speech Recognition evaluation }
The above experimental results indicate that the speech synthesized by SVD-modified GhostVec has a unique speaking voice and an acceptable difference from genuine speech audio. These two conclusions motivate us to explore the practical implications of our synthetic audio. We mainly verify from the perspective of ASR.

Firstly, we observe the Mel-spectrum of different embeddings. Figure \ref{fig:mel} shows the spectrum from different genders and various speaker embeddings. The results show that some adversarial elements in the spectrum still influence the length and position of silent frames. In objective evaluation on section \ref{objection}, we found that the synthesized audios contained sufficient target speaker information. Therefore, we convey those audios to the ASR systems. The objective intelligibility evaluation in terms of CER in Table \ref{tab:cer} shows the quality of the speech content. We synthesize the same text as the baseline and get similar scores. The difference in scores may be due to the residual adversarial elements. That is exactly what is so special about GhostVec's synthetic audio. The experimental results also directly reflect that the audio synthesized in this paper can be used as valid audio. It is possible to improve the model's robustness further.

\section{Conclusions}
This paper shows a new threat to speaker privacy from an End-to-End ASR system. In this paper, we use adversarial examples to extract a target speaker embedding, GhostVec, from the ASR model without using any reference speech. The target speaker's voice is further synthesized using SVD-modified GhostVec. The experimental results show the effectiveness of our method in terms of both the synthesized audio quality and the speaker characteristics in audio. Moreover, the intelligibility of synthesized audio also confirms that audio synthesized by our proposed model can serve as legitimate audio samples. We hope the discovery in this study will catalyze future downstream research on speaker and speech privacy preservation topics.


%\newpage

\bibliographystyle{ACM-Reference-Format}
\bibliography{mybib}


\end{document}